\documentclass[10pt,letterpaper]{article}
\usepackage[top=0.85in,left=2.75in,footskip=0.75in]{geometry}

\usepackage{amsmath,amssymb}

\usepackage{changepage}

\usepackage[utf8x]{inputenc}

\usepackage{textcomp,marvosym}

\usepackage{cite}

\usepackage{nameref, hyperref}

\usepackage[right]{lineno}

\usepackage{microtype}
\DisableLigatures[f]{encoding = *, family = * }

\usepackage[table]{xcolor}

\usepackage{array}

\newcolumntype{+}{!{\vrule width 2pt}}

\newlength\savedwidth

\raggedright
\usepackage[aboveskip=1pt,labelfont=bf,labelsep=period,justification=raggedright,singlelinecheck=off]{caption}

\makeatletter
\renewcommand{\@biblabel}[1]{\quad#1.}
\makeatother

\usepackage{lastpage,fancyhdr,graphicx}
\usepackage{epstopdf}
\fancyheadoffset[L]{2.25in}
\fancyfootoffset[L]{2.25in}
\usepackage[T1]{fontenc}

\usepackage{xspace}

\usepackage{booktabs}

\usepackage{amsthm}
\newtheorem*{definition}{Definition}
\newtheorem*{theorem}{Theorem}

\newcommand{\algorithm}[1]{\textsf{#1}\xspace}

\newcommand{\problem}[1]{\textsc{#1}\xspace}

\newcommand{\netcon}{\algorithm{netcon}}
\newcommand{\freetdi}{\algorithm{freetdi}}
\newcommand{\meijie}{\algorithm{meiji-e}}

\newcommand{\quickbb}{\algorithm{quickbb}}

\newcommand{\maxcut}{\problem{MaxCut}}

\newcommand{\tw}{$tw$\xspace}
\newcommand{\cc}{$cc$\xspace}

\newcommand{\software}{\algorithm{ConSequences}}
\newcommand{\github}{\url{github.com/TheoryInPractice/ConSequences}}
\newcommand{\liquid}{LIQUi|>\xspace}

\newcommand{\backgroundsection}{Background\xspace}
\newcommand{\codesection}{\algorithm{ConSequences}: An Accessible, Extendable Framework\xspace}
\newcommand{\merasection}{MERA Applications\xspace}
\newcommand{\qtorchsection}{Applications with qTorch Simulator\xspace}

\definecolor{blue_background}{rgb}{0.65098039215686276, 0.80784313725490198, 0.8901960784313725}
\definecolor{blue}{rgb}{0.12156862745098039, 0.47058823529411764, 0.70588235294117652}
\definecolor{green_background}{rgb}{0.69803921568627447, 0.87450980392156863, 0.54117647058823526}
\definecolor{green}{rgb}{0.20000000000000001, 0.62745098039215685, 0.17254901960784313}
\definecolor{red_background}{rgb}{0.98431372549019602, 0.60392156862745094, 0.59999999999999998}
\definecolor{red}{rgb}{0.8901960784313725, 0.10196078431372549, 0.10980392156862745}

\usepackage{multicol}
\def\boxitblue#1{%
  \smash{\color{blue}\fboxrule=1pt\relax\fboxsep=2pt\relax%
  \llap{\rlap{\fbox{\vphantom{0}\makebox[#1]{}}}~}}\ignorespaces
}

\def\boxitred#1{%
  \smash{\color{blue_background}\fboxrule=1pt\relax\fboxsep=2pt\relax%
  \llap{\rlap{\fbox{\vphantom{0}\makebox[#1]{}}}~}}\ignorespaces
}

\usepackage[framemethod=tikz]{mdframed}

\newmdenv[innerlinewidth=1pt, roundcorner=4pt,linecolor=blue]{myboxblue}

\newmdenv[innerlinewidth=1pt, roundcorner=4pt,linecolor=blue_background]{myboxred}

\begin{document}
\vspace*{0.2in}

\begin{flushleft}
{\Large
\textbf\newline{Benchmarking treewidth as a practical component of tensor-network--based quantum simulation} 
}
\newline
\\
Eugene F. Dumitrescu\textsuperscript{1, \dag},
Allison L. Fisher\textsuperscript{2},
Timothy D. Goodrich\textsuperscript{2, *},
Travis S. Humble\textsuperscript{1, \dag},
Blair D. Sullivan\textsuperscript{2},
Andrew L. Wright\textsuperscript{2}
\\
\bigskip
\textbf{1} Quantum Computing Institute, Oak Ridge National Laboratory, Oak Ridge, TN, USA
\\
\textbf{2} Department of Computer Science, North Carolina State University, Raleigh, NC, USA
\\
\bigskip
\dag This manuscript has been authored by UT-Battelle, LLC under Contract No. DE-AC05-00OR22725 with the U.S. Department of Energy. The United States Government retains and the publisher, by accepting the article for publication, acknowledges that the United States Government retains a non-exclusive, paid-up, irrevocable, world-wide license to publish or reproduce the published form of this manuscript, or allow others to do so, for United States Government purposes. The Department of Energy will provide public access to these results of federally sponsored research in accordance with the DOE Public Access Plan (http://energy.gov/downloads/doe-public-access-plan).\\

* Corresponding author: tdgoodri@ncsu.edu

%
%
%

%
%
%

\end{flushleft}
\section*{Abstract}
\label{section:abstract}
Tensor networks are powerful factorization techniques which reduce resource requirements for numerically simulating principal quantum many-body systems and algorithms.
The computational complexity of a tensor network simulation depends on the tensor ranks and the order in which they are contracted. Unfortunately, computing optimal contraction sequences (orderings) in general is known to be a computationally difficult (NP-complete) task. In 2005, Markov and Shi showed that optimal contraction sequences correspond to optimal (minimum width) tree decompositions of a tensor network's line graph, relating the contraction sequence problem to a rich literature in structural graph theory.
While treewidth-based methods have largely been ignored in favor of dataset-specific algorithms in the prior tensor networks literature, we demonstrate their practical relevance for problems arising from two distinct methods used in quantum simulation: multi-scale entanglement renormalization ansatz (MERA) datasets and quantum circuits generated by the quantum approximate optimization algorithm (QAOA).
We exhibit multiple regimes where treewidth-based algorithms outperform domain-specific algorithms, while demonstrating that the optimal choice of algorithm has a complex dependence on the network density, expected contraction complexity, and user run time requirements. We further provide an open source software framework designed with an emphasis on accessibility and extendability, enabling replicable experimental evaluations and future exploration of competing methods by practitioners.


\section*{Introduction}
\label{section:introduction}
Tensor network factorizations provide a framework for controlled approximation which exponentially reduces the memory required to simulate a variety of quantum many-body systems~\cite{white1992density, vidal2006class} and circuits~\cite{vidal2003efficient, markov2008simulating}.
These factorizations do so by representing targeted sub-sectors of the full (exponentially scaling) Hilbert space.
The tensors comprising the factorization are placed on the vertices of a graph, one appropriate to the geometry under consideration, and are contracted along the edges as needed to compute physical observables~\cite{white1992density}.

Since their early usage as the density matrix renormalization group description for gapped spin chains~\cite{white1992density, biamonte2017tensor}, tensor networks have been adapted and reformulated to also describe 2D area-law states~\cite{eisert2010area, schuch2011classifying},
critical systems \cite{vidal2006class}, lattice gauge theories~\cite{rico2014tensor, pichler2016real}, AdS/CFT duality~\cite{pastawski2015holographic},
and open quantum systems~\cite{verstraete2004matrix, werner2016positive}.
In addition to describing a wide range of physical phenomena, satisfiability problems~\cite{biamonte2015tensor,biamonte2017tensor} and quantum computing simulations~\cite{vidal2003efficient, markov2008simulating} can be formulated as tensor contraction problems.
In the case of the latter, simulations of quantum error correcting codes are leading to important insights into fault tolerant quantum computation~\cite{ferris2014tensor, darmawan2017tensor}.
Additionally, given the tremendous interest in validating increasing complex experimental quantum computations~\cite{boixo2018characterizing},
a flurry of simulations have recently appeared in which the underlying tensor network graph emerges from the structure of the algorithm being employed~\cite{dumitrescu2017tree, dang2017optimising, fried2017qtorch, pednault2017breaking}.

The overall descriptive power and algorithmic computational complexity, as formalized by the \emph{contraction complexity} of a tensor network~\cite{markov2008simulating}, is determined by the tensor network construction's underlying graph structure.
For some tensor network algorithms (e.g. the matrix product state formulation) the contraction complexity is fixed and well understood.
However, the task of determining the contraction complexity in general, along with computing an optimal contraction sequence witnessing this complexity, is NP-complete~\cite{chi1997optimizing}.

Despite this daunting theoretical complexity, efficient methods exist in practice for obtaining both optimal and `good enough' contraction sequences.
Domain-specific approaches typically search the space of all possible sequences and apply heuristic pruning techniques to reduce the search space~\cite{chi1997optimizing, pfeifer2014faster, fried2017qtorch}.
Effective algorithms in this area incorporate pruning rules proprietary to the target application's data (e.g., MERA networks~\cite{pfeifer2014faster}), which limits their broader applicability.
Another standard technique involves transforming the tensor network into a line graph, then computing a perfect elimination ordering and its treewidth, which can be translated into a contraction sequence and complexity for the original network, respectively~\cite{markov2008simulating}.

In practice, however, engineering issues prevent these methods from being applied effectively.
Domain-specific approaches typically suffer from proprietary construction, each assuming a different representation of the tensor networks, and using different code languages, dependencies, and interfaces (often with little-to-no documentation).
Additionally, these implementations are typically only tested on the data for which they were designed, providing no expectation for how they might perform and/or scale in different contexts.
Treewidth-based approaches further suffer from the graph theory overhead needed to convert their (typical) output of tree decompositions into perfect elimination orderings for the line graph and then contraction sequences for the tensor network.

Our primary contribution to the literature is to provide an open source code framework (available at \github) for integrating all existing contraction sequence algorithms into a common interface designed for extendability and documented for accessibility; further, we have tabulated the performance of several leading contraction sequence algorithms. Our results provide quantum circuit simulation developers an extended benchmark for expected performance on circuits with varying structures and complexities.

We use container-based (Docker~\cite{merkel2014docker}) wrappers for each contraction sequence algorithm, completely removing code dependency issues, and provide Python-based utilities for converting various input/output formats into standardized internal formats for consistency.
We demonstrate the utility of this software by reproducing two previous studies based on domain-specific algorithms, and extending them to include treewidth-based solvers in a broader set of experimental results.
We find that modern treewidth solvers from the recent PACE 2017 coding challenge~\cite{dell2017pace} are both faster and have more consistent run times than the domain-specific algorithms.
This speed increase allows us to study larger datasets in both experiments, and provide more competitive comparisons of a tensor network simulator~\cite{fried2017qtorch} against Microsoft's \liquid Hilbert space simulator~\cite{wecker2014liquid}.
In particular, we show that contraction sequence algorithms are no longer the major bottleneck in tensor network simulations, and there is immediate value in work improving the scalability of downstream contraction code.

The paper is organized as follows. We begin with relevant definitions and an overview of related work in the {\backgroundsection}, then
describe the functionality of our code framework and considerations for use and extension in {\codesection}.
In the subsequent {\merasection} section, we reproduce a study by Pfeifer et al.~\cite{pfeifer2014faster}, evaluating their algorithm \netcon alongside two treewidth algorithms from the PACE 2017 challenge (\freetdi~\cite{freetdi} and \meijie~\cite{meiji}) on a dataset including multi-scale entanglement renormalization ansatz (MERA) networks~\cite{vidal2006class}. We extend this initial comparison on a larger corpus of MERA networks, pushing the limits of these exact contraction sequence solvers on a new benchmark.
In the {\qtorchsection} section, we reproduce a study by Fried et al.~\cite{fried2017qtorch}, evaluating another treewidth-based solver (\quickbb~\cite{gogate2004complete}) against \freetdi and \meijie on quantum circuits formulated with Farhi et al.'s quantum approximate optimization algorithm (QAOA) for \maxcut on $r$-regular graphs~\cite{farhi2014quantum}.
In addition to contraction sequence comparisons, we simulate the tensor network with qTorch~\cite{fried2017qtorch}, noting the correlation between simulation time and contraction complexity, and providing an updated comparison with Microsoft's \liquid simulator~\cite{wecker2014liquid}.
We conclude with a summary and directions for future work.

\section*{\backgroundsection}
\label{section:background}
For a graph $G$, we use $V(G)$ and $E(G)$ to denote the sets of vertices and edges, respectively, and use $G[X]$ to denote the subgraph induced by a set of vertices $X$. We say two vertices $u,v$ are \emph{adjacent} if $(u,v) \in E(G)$, and call the set of all vertices adjacent to $v$ its \emph{neighborhood} $N(v)$.
The \emph{degree} of $v$ is $|N(v)|$, and a graph is $r$-regular if every vertex has degree $r$.

Formally, a \emph{tensor network} is represented by a graph whose vertices correspond to tensors and edges denote
tensor contractions over tensor indices. A \emph{contraction} of two tensors corresponds to an \emph{edge contraction} in the graph, where two vertices $u, v$ with respective neighborhoods $N(u), N(v)$ are replaced with a single vertex $uv$ with neighborhood $(N(u) \cup N(v))\setminus \{u,v\}$.

In the remainder of this section we define contraction complexity and its relationship to notions from structural graph theory including treewidth, then outline the methods used to generate MERA and QAOA tensor networks, which are used as data in our experiments.

\subsection*{Contraction Complexity and Treewidth}
Simulation of a tensor network requires its contraction down to a single tensor, and the network's structure imposes certain lower bounds on the information that must be kept in the network, fundamentally captured in the notion of contraction complexity:

\begin{definition}[Contraction Complexity (\cc)]
A \emph{contraction sequence} is an ordering of a tensor network's edges, and the \emph{complexity} of a contraction sequence $S$ is the largest degree of a merged vertex created by contracting the network according to $S$.
The \emph{contraction complexity} (\cc) of a tensor network is the minimum complexity over all possible contraction sequences.
\end{definition}

Computing the contraction complexity optimally is an NP-hard~\cite{chi1997optimizing} optimization problem with strong ties to structural graph theory under the guise of treewidth and elimination orderings.

\begin{definition}[Treewidth (\tw)]
A \emph{tree decomposition} of a graph $G$ is a tree $T$ with a function $f$ mapping nodes in $T$ to \emph{bags} (sets) of vertices from $G$, such that the following conditions hold:
\begin{enumerate}
\item All vertices are represented: $\bigcup_{t \in V(T)} f(t) = V(G)$.
\item All edges are represented: $\forall (u, v) \in E(G), \exists t \in V(T)
\text{ s.t. } u, v \in f(t)$.
\item Graph vertices induce a (connected) subtree of $T$: if $w \in f(r) \cap f(s)$ for $r,s \in V(T), w \in V(G)$, then $w \in f(t)$  for all $t$ on the path from $r$ to $s$ in $T$.
\end{enumerate}
The \emph{width} of a tree decomposition is $\max_{t \in V(T)} |f(t)| - 1$, and the \emph{treewidth} of a graph $G$, denoted $tw(G)$, is the minimum width over all valid tree decompositions of $G$.
\end{definition}

Perhaps surprisingly, treewidth can be viewed as a vertex-centric formulation of the edge-centric contraction complexity, via a transformation of the underlying graph $G$ to its \emph{line graph} $L(G)$.
The line graph is constructed with $V(L(G)) = E(G)$ and $E(L(G)) = \{(e_1, e_2) ~|~ e_1 \neq e_2 \in E(G)$ s.t. $e_1, e_2$ share a common endpoint$\}$.
The treewidth of $L(G)$ then captures the same complexity:

\begin{theorem}[Markov and Shi \cite{markov2008simulating}]
The contraction complexity of a graph equals the treewidth of its line graph.
\end{theorem}

The relationship between treewidth and contraction sequences is perhaps easier seen through
the characterization of treewidth using \emph{elimination orderings}, which are permutations of the vertices. Given an
elimination order $\pi =  v_1, v_2, \ldots, v_n$ of a graph G, the \emph{fill-in graph} $G_\pi$ is constructed by iterating over $v_i$ from $i = 1$ to $n$
and adding edges to make the neighbors of $v_i$ in $G[v_{i}, \ldots, v_n]$ a clique.
A graph has treewidth at most $k$ if and only if there exist an elimination ordering $\pi$ so that each vertex has at most $k$ higher numbered neighbors in $G_\pi$ (see e.g., ~\cite{bodlaender1998treewidth} for a proof). This condition naturally corresponds to the maximum degree of a tensor in the contraction sequence.

While this intuition provides an straightforward mapping from a contraction sequence to a tree decomposition, the other direction of Markov and Shi's proof shows how to convert an arbitrary tree decomposition into a contraction sequence with equal complexity.
This non-trivial conversion allows treewidth solvers, whose native outputs are only tree decompositions, to be used directly as contraction sequence algorithms.
To enable future work, we provide an implementation of this conversion as a modular subroutine in our post-processing utilities.

Rapid advances have been made in treewidth solvers in recent years, in large part to the Parameterized Algorithms and Computational Experiments (PACE) Challenge ~\cite{dell2016first, dell2017pace}.
Previous algorithms with practical implementations (such as \quickbb~\cite{gogate2004complete}) are based on searching the space of elimination orderings and given the connection between contraction sequences and elimination orderings~\cite{markov2008simulating}, share a strong resemblance to typical domain-specific algorithms~\cite{chi1997optimizing, pfeifer2014faster}.

However, recent work in separator-based treewidth algorithms has begun to dominate modern benchmarks.
The classic Arnborg, Corniel, and Proskurowski dynamic programming algorithm~\cite{arnborg1987complexity}, reformulated as a positive-instance dynamic programming (PID) algorithm, has produced the winners of both the PACE 2016 (a Java implementation by Tamaki) and PACE 2017 (a C++ implementation by Larisch and Salfelder, \freetdi) challenges~\cite{dell2016first, dell2017pace, freetdi}.
The 2017 challenge also saw a better scaling implementation (\meijie~\cite{meiji}) based off of a PID-reformulation of the improved dynamic programming algorithm by Bouchitt{\'e} and Todinca~\cite{bouchitte2001treewidth}.

\subsection*{MERA Tensor Networks}
\begin{figure}[!ht]
\begin{adjustwidth}{-2.25in}{0in}
\centering
\includegraphics[width=0.55\linewidth]{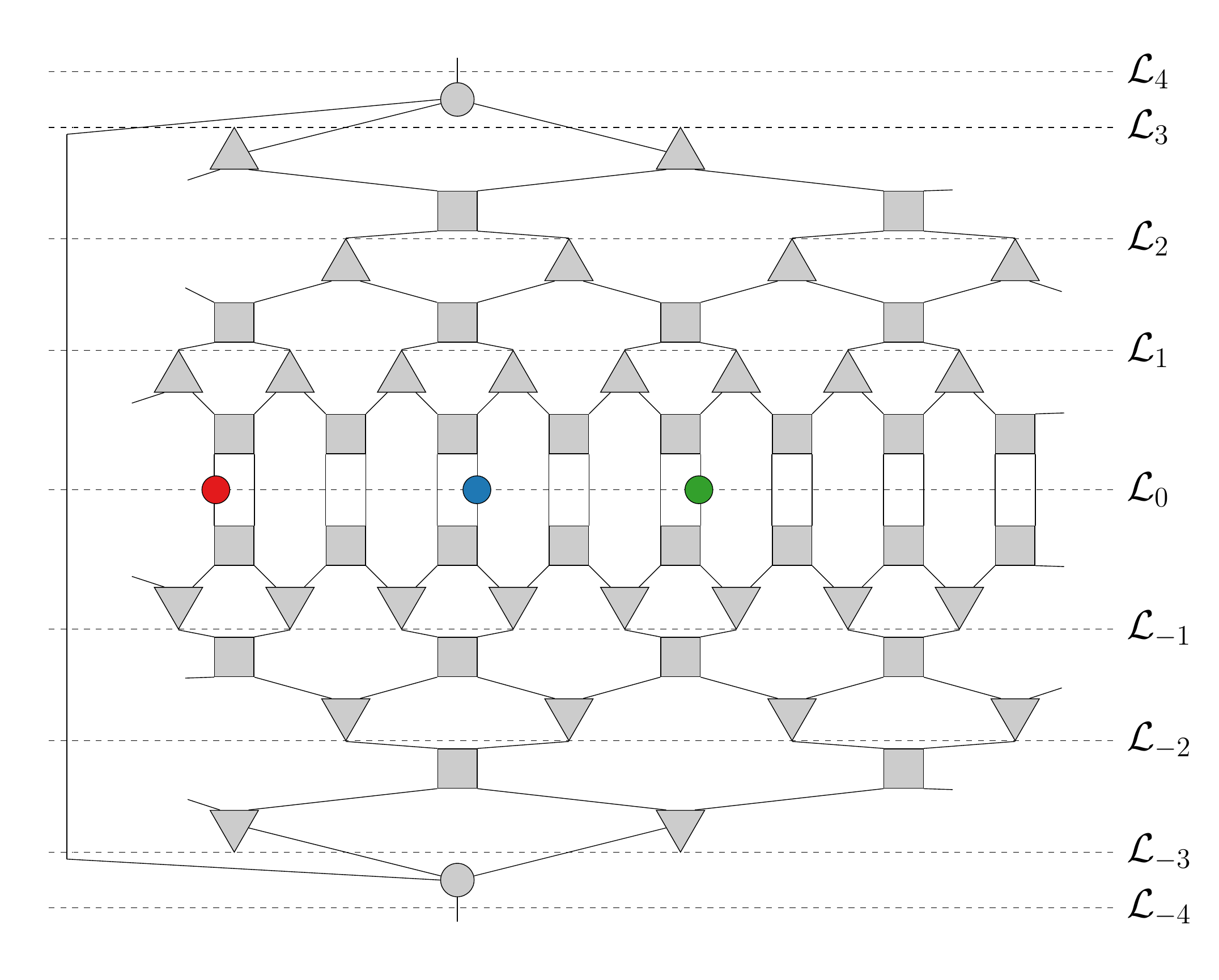}\\
\includegraphics[width=0.3\linewidth]{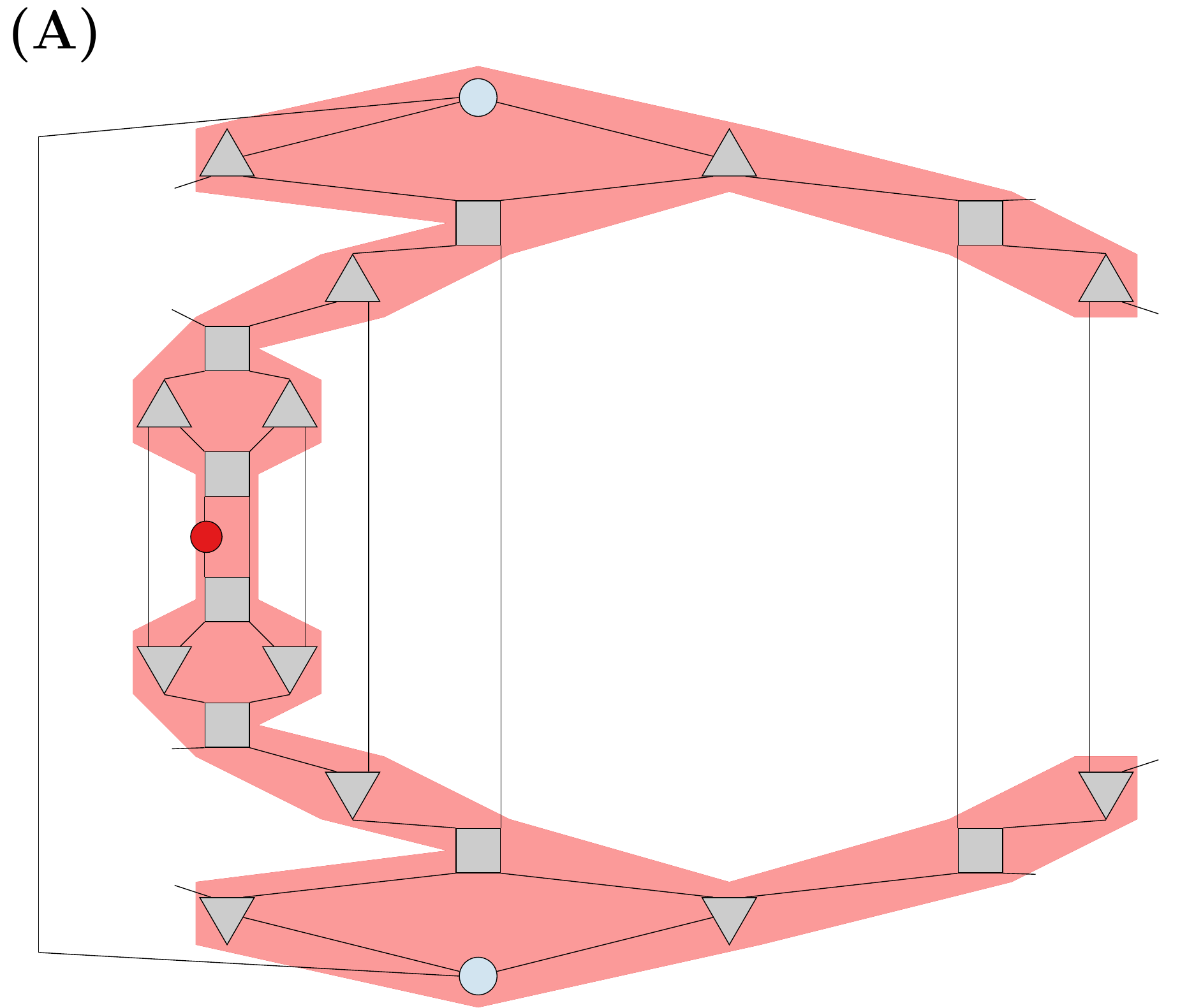}
\includegraphics[width=0.3\linewidth]{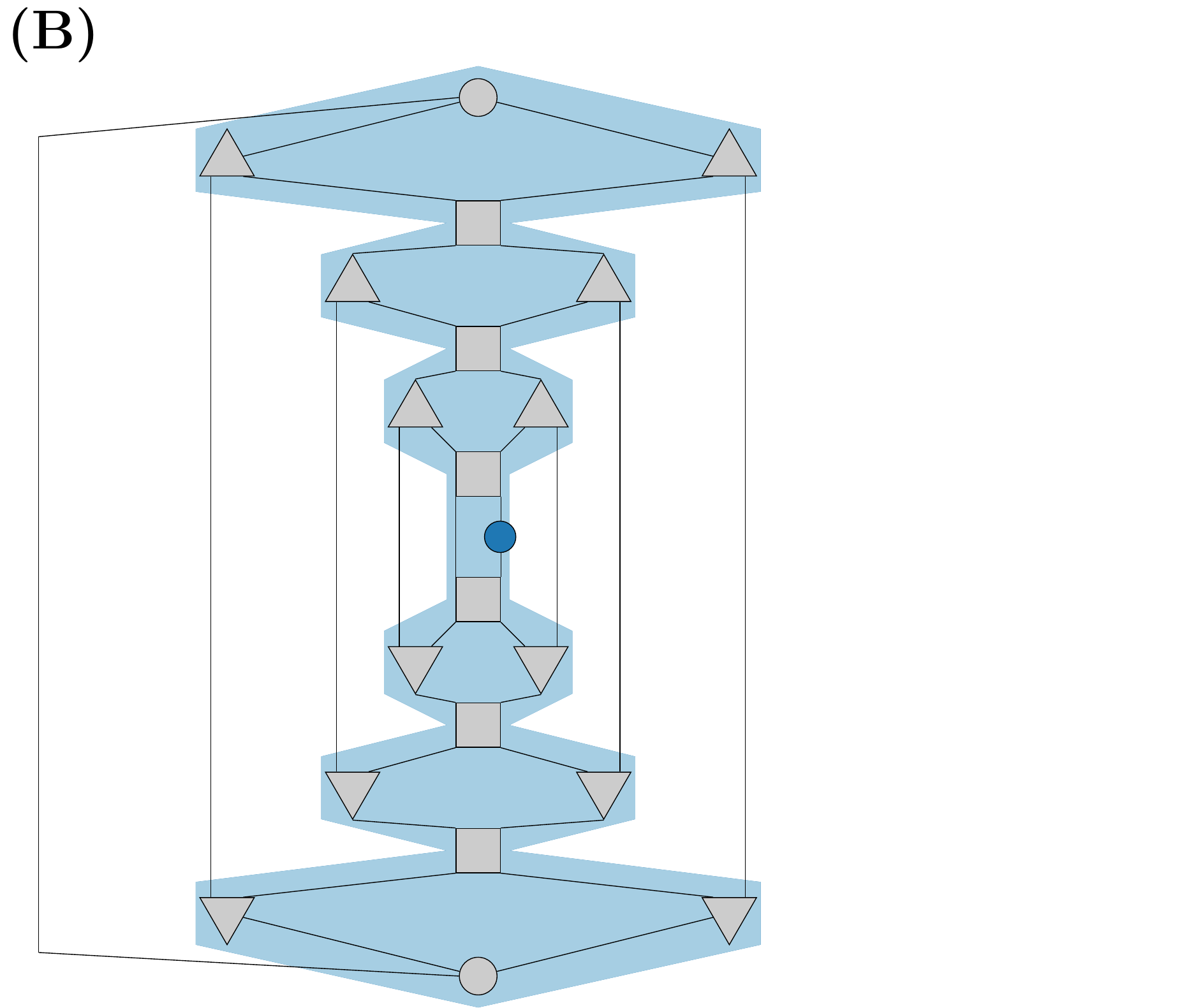}
\hspace{-.75in}
\includegraphics[width=0.3\linewidth]{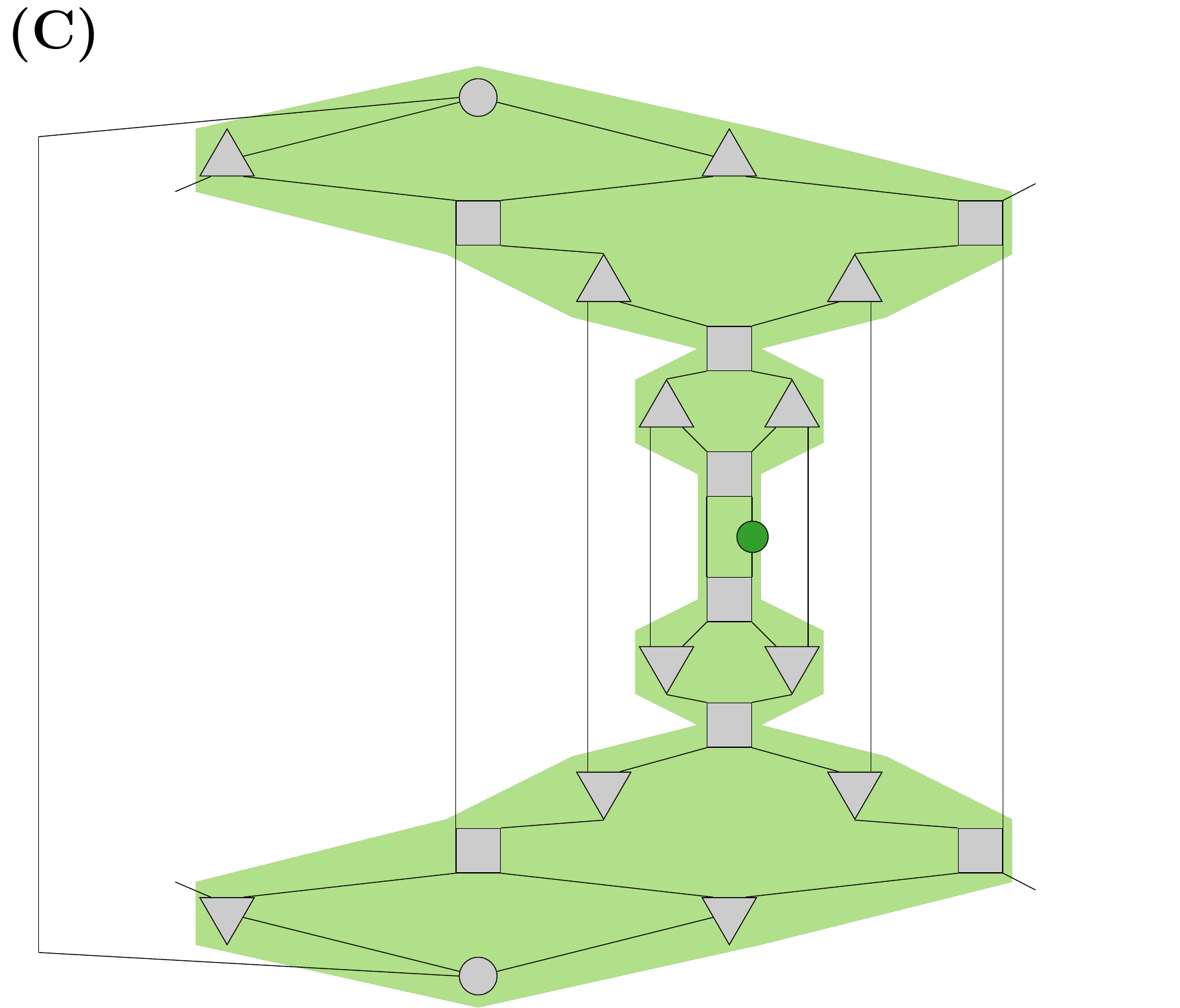}
\caption{(Top) A 1D binary MERA with a 16-site lattice and 3 levels of coarsening; three operator placements are highlighted (red, blue, green). (Bottom) Causal cones and final tensor networks for each of the three highlighted operators. Note that the tensor networks for the left-most (red) and right-most (green) operators are isomorphic to one another, but structurally distinct from the middle (blue) operator's network.}
\label{figure:mera}
\end{adjustwidth}
\end{figure}

One class of tensor networks that we examine is the multi-scale entanglement renormalization ansatz (MERA).
Given that contraction sequence algorithms only utilize the structure of the underlying graph in these networks, we restrict our presentation here to the important structural notions (visualized with a 1D binary MERA in Fig~\ref{figure:mera}).
We direct the interested reader to \cite{evenbly2009algorithms} for a rigorous description beyond the graph structure.

Fundamentally, MERA is a scheme for mapping a lattice of \emph{operator sites} onto a coarser lattice.
This mapping is expressed in terms of \emph{coarsening layers}.
The lattices on which MERA acts have an inherent dimension, which we denote $d$; for simplicity we only consider examples in 1- and 2-dimensions in this paper.
The most detailed lattice ($\mathcal{L}_0$) contains all sites for operators, and lattice $\mathcal{L}_1$ is produced after one level of coarsening.
A coarsening level consists of a layer of unitaries followed by a layer of isometries.
To disentangle the sites, in the MERAs we consider, \emph{unitary} tensors take in $2^d$ wires (edges) and output $2^d$ wires for a lattice of dimension $d$.
For the coarsening layer, \emph{$k:1$ isometry} tensors take in $k$ wires and output one wire.
In total, going from lattice $\mathcal{L}_i$ with $s$ sites to $\mathcal{L}_{i+1}$ requires $\frac{s}{2}$ unitary tensors, $\frac{s}{2}$ isometry tensors, and produces a new lattice with $\frac{s}{k}$ sites.
This structure is then reflected for negative lattice levels, and a wire connects the top and bottom level interface tensors.
Once this MERA graph is defined, operators are placed on lattice sites in $\mathcal{L}_0$ and the \emph{causal cone} is computed by including the operators and any tensor that lies on an ascending (descending) path to the upper (lower) interface tensor (Fig~\ref{figure:mera}).
Once the causal cone is computed, every tensor not included in the cone is removed (by unitarity and properties of the isometries), and wires are added from a tensor to its dual mirror such that all tensors have the requisite number of wires.

\subsection*{QAOA Quantum Circuits}
Another source for data comes from the quantum approximation optimization algorithm (QAOA)~\cite{farhi2014quantum}, a hybrid classical-quantum algorithm for utilizing near-term ($\sim$ 100 qubit) quantum computers.
While applicable to generic satisfiability problems, we restrict ourselves to the \maxcut optimization problem on $r$-regular graphs.
Notably, when QAOA is applied to the \maxcut problem, the structure of the input graph is reflected in the quantum circuit.
Interested readers should refer to Farhi et al.'s formulation of QAOA~\cite{farhi2014quantum} for a theoretical treatment, or the qTorch source code~\cite{fried2017qtorch} for a practical example.

\section*{\codesection}
\label{section:contraction-suite}
One issue preventing widespread experimentation with (and adoption of) contraction sequence algorithms was the practical problem of installing the software and managing software dependencies.
Of the algorithms presented in this paper, one is interfaced with MATLAB and uses C extensions for computationally-difficult sections (\netcon), one is written in Java (\meijie), two are written in C++ (\freetdi, qTorch), and one is \emph{only} distributed as a binary executable for Linux (\quickbb).
Often these implementations were written as a proof of concept and contain little-to-no documentation, especially regarding the code library dependencies needed to compile the code.

Additionally, once the code is compiled, each solver has proprietary input and output format.
Algorithms from the treewidth literature may require the input graph to have particular vertex labels, and typically output a tree decomposition or an elimination ordering.
Algorithms from the contraction sequence literature may require the input as a quantum circuit in Quantum Assembly format or as a tensor network, and the contraction sequence output may be an ordering of edges in the network or a sequence of contractions that automatically removes resulting self-loops.

\begin{figure}[!ht]
\centering
\includegraphics[width=\linewidth]{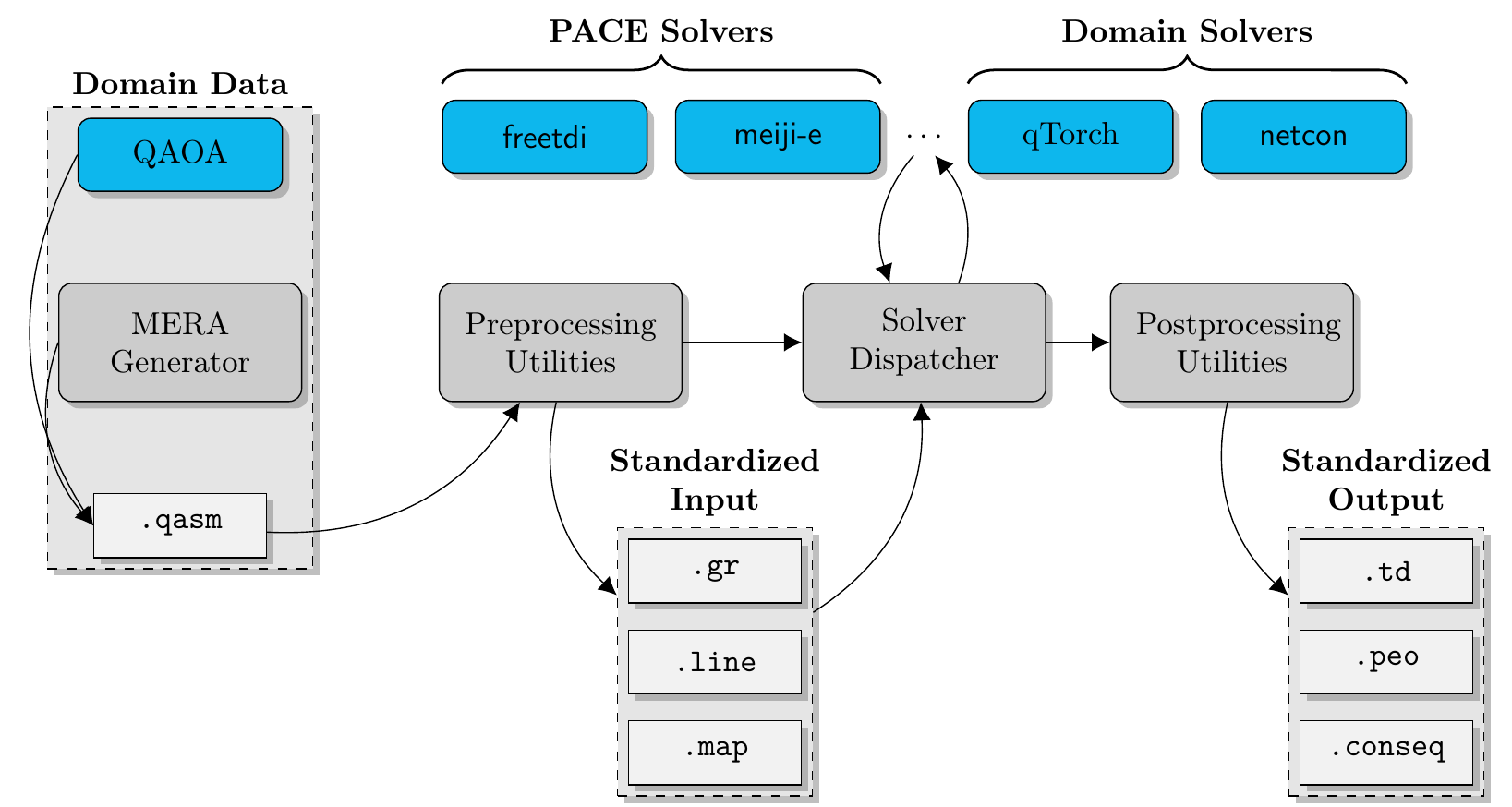}
\caption{Visualization of the \software pipeline. Externally-generated domain data is parsed into standardized graph formats with the pre-processing utility. The solver dispatcher then allows the user to compute contraction sequences (or their equivalent, e.g. tree decompositions) using external algorithms in Docker containers, then executes a post-processing utility to output standardized formulations of a contraction sequence.}
\label{figure:code}
\end{figure}

Addressing both problems at once, we provide an open source framework \software (Fig~\ref{figure:code}) for running contraction sequence algorithms, designed to be both \emph{accessible} and \emph{extendable}.
The code is available at \github, complete with documentation for using existing code and tutorials for extending the functionality.
At the core of this framework is a pipeline for pre-processing input, dispatching solvers, and post-processing output.
The pre-processing utilities take tensor networks in formats such as Quantum Assembly and various graph formats, then generate graph files in a standardized format.
The solver dispatcher manages calls to contraction sequence algorithms, and includes functionality such as parallelism, managing seeds, and handling timeouts.
The post-processing utilities then take output from these solvers and generate a full set of output (including tree decompositions, perfect elimination orderings, and contraction sequences), allowing various aspects of the output to be reported and analyzed.

To make the framework \emph{accessible}, the central pipeline is implemented as lightweight Python files, and individual contraction sequence algorithms are wrapped inside of Docker~\cite{merkel2014docker} images.
Docker images are similar to virtual machines in that they abstract away all dependency issues, but have the distinct advantage of very little CPU and memory footprint on a native Linux system.
Users need only install Python and Docker to run the code on Windows, MacOS or Linux; these steps are especially straightforward on Linux and we provide instructions for new users.

To make the framework \emph{extendable}, we provide Docker templates for data generators (e.g., MERA networks) and contraction sequence algorithms.
These templates are accompanied by tutorials which guide new users through wrapping up their code in Docker and interacting with our existing structure properly.

All experiments were run on three identical workstations, each with a single Xeon E5-2623 v3 processor (8 threads with a 3.0GHz base clock and 10MB cache) and 64GB system memory.
Contraction sequence algorithms were run with a single thread and \liquid and qTorch simulations with all threads.
Experiments were run one at a time on the workstations, preventing noise from non-uniform cache usage between competing jobs.
These workstations ran Fedora 27 with Docker 18.03.1-ce and Python 3.6.5. Algorithm-dependent software requirements (e.g., gcc, MATLAB) are fixed per algorithm in its Docker image wrapper; details are deferred to the code repository.

\section*{\merasection}
\label{section:applications-mera}
In this section we compare exact treewidth solvers to Pfeifer et al.'s \netcon algorithm~\cite{pfeifer2014faster}, on data from MERA networks.
A small corpus of datasets from~\cite{pfeifer2014faster} is initially considered, in which case optimal contraction sequences are found within four seconds by both treewidth solvers.
To analyze algorithmic scaling with an extended benchmark, we generate 1- and 2-dimensional MERA networks with all possible placements of 1- and 2-operators, where we additionally observe that the \meijie treewidth solver scales better than \freetdi when networks become dense and increase in contraction complexity.

\subsection*{Initial Comparison on Pfeifer Benchmark}
The \netcon implementation was developed as the contraction sequence algorithm for a simulation toolset written for MATLAB \cite{pfeifer2014ncon, evenbly2014improving, pfeifer2014faster}.
In the vein of previous approaches~\cite{chi1997optimizing} such as depth-first search and dynamic programming, \netcon's core subroutine is a breadth-first search (BFS) over the solution space of all possible contraction sequences.
To trim down this exponentially-sized space, the authors introduce two pruning methods for reducing the search space at each step of the BFS:
first, if a contraction would cost more than a user-defined threshold, this contraction will not be considered;
second, the authors provide a list of criteria for when outer product contractions should not be made.
This algorithm is exact (i.e., it finds optimal contraction complexity), but its run time depends heuristically on the effectiveness of the pruning techniques to a particular network's search space.
Provided as a MATLAB package, the core subroutines in \netcon are implemented in external C code for efficiency.
The authors of~\cite{pfeifer2014faster} evaluate their pruning heuristics on seven networks, including Tree Tensor Networks (TTN), Time-Evolution Block Decimation (TEBD), and MERA networks ranging from five to 27 tensors.
In this previous work, the authors found that the the pruning techniques resulted in faster results on all networks, with the largest network requiring 36 seconds.

\begin{table}[!ht]
\tabcolsep = 1.5mm
\centering
\begin{tabular}{lrrrp{0.1in}rrrr}
\toprule
\multicolumn{4}{c}{\textbf{Instance}}
& & \multicolumn{3}{c}{\textbf{Run Time (sec)}}\\
\cmidrule(lr){1-4}
\cmidrule(lr){6-8}
Name & $|V|$ & $|E|$ & \cc && \freetdi & \meijie & \netcon \\
\midrule

{3:1 1D TTN} & {5} & {28} & {6} && {0.009} & {0.338} & {0.002}\\
{TEBD} & {6} & {24} & {3} && {0.002} & {0.297} & {0.002}\\
{3:1 1D MERA} & {7} & {45} & {8} && {0.005} & {0.381} & {0.007}\\
{9:1 2D TTN} & {9} & {38} & {12} && {0.004} & {0.459} & {0.004}\\
{2:1 1D MERA} & {11} & {34} & {9} && {0.005} & {0.386} & {0.016}\\
{9:1 2D MERA} & {19} & {62} & {16} && {0.011} & {0.669} & {0.051}\\
{4:1 2D MERA} & {27} & {55} & {26} && {2.944} & {3.534} & {24.415}\\

\bottomrule
\end{tabular}

\caption{Run times for each contraction sequence algorithm when executed on tensor network datasets from Pfeifer et al. \cite{pfeifer2014faster}. For each tensor network, the number of tensors ($|V|$), edges ($|E|$), and optimal contraction complexity (\cc) are reported.}
\label{table:mera_preliminary}
\end{table}

We reproduce this experiment on all seven networks (provided in~\cite{pfeifer2014faster}).
For \netcon we use the MATLAB interface with an external C package, using all optimizations as specified in the accompanying code~\cite{pfeifer2014faster}.
We compare against two exact PACE algorithms, \freetdi and \meijie.
Refer to the \software section for workstation specifications.
As seen in Table~\ref{table:mera_preliminary}, both \netcon and \freetdi require on the order of 0.001 seconds on the first four networks, whereas \meijie was two orders of magnitude slower.
On the three larger networks, however, \freetdi was fastest, with both PACE algorithms finishing within four seconds on the largest network.
This last data point is of particular interest because it hints at scalability differences between these algorithms.
Whereas \meijie's run time was $5\times$ slower on the 4:1 2D MERA compared to the 9:1 2D MERA, \freetdi increased to over $250\times$ slower and \netcon over $450\times$ slower.

\subsection*{Extended Benchmark on Large MERA Networks}

To test scalability further, we compared the algorithms' performance on a larger corpus of MERA networks.
As seen in Fig~\ref{figure:mera}, the iterative layers of isometries and unitaries in MERA networks allow one to easily generate underlying graphs given a unitary and isometry specification.
We provide a MERA generator in our code for 1D lattices with binary isometries and 2D lattices with 4:1 isometries.
This generator takes as input the number of coarsening levels and whether the top level of isometries should connect to a common tensor in the style of \cite{evenbly2009algorithms}.
Once a MERA network is generated from these parameters, the locations of operators to be evaluated are chosen, a causal cone is computed, and the network is reduced down to the final tensor network by simplifying the network outside the causal cone.

\begin{table}[!hb]
\tabcolsep = 1.5mm
\centering
\begin{multicols}{3}

\begin{myboxblue}
\footnotesize
\centering
\begin{tabular}{rrrr}
\multicolumn{4}{c}{\textbf{1D 2:1 MERA}}\\
\toprule
\multicolumn{4}{c}{\textbf{$\ell = 6$, 2 oper.}}\\
\midrule
$|V|$ & $|E|$ &  $|{\mathcal{S}}|$ & $|\widehat{\mathcal{S}}|$ \\
\midrule
40 & 65 & 1 & 1 \\
62 & 104 & 2 & 1 \\
64 & 107 & 2 & 1 \\
66 & 110 & 2 & 1 \\
68 & 113 & 2 & 1 \\
68 & 114 & 2 & 1 \\
70 & 116 & 2 & 1 \\
70 & 117 & 2 & 1 \\
72 & 120 & 2 & 1 \\
74 & 123 & 2 & 1 \\
74 & 124 & 4 & 3 \\
76 & 127 & 4 & 3 \\
78 & 130 & 4 & 3 \\
80 & 134 & 8 & 7 \\
82 & 137 & 8 & 7 \\
86 & 144 & 16 & 15 \\
\bottomrule
\end{tabular}
\end{myboxblue}

\begin{tabular}{rrrr}
\toprule
$\ell$ & $|V|$ & $|\mathcal{M}|$ & $|\widehat{\mathcal{M}}|$ \\
\midrule
\multicolumn{4}{l}{\textbf{1D 2:1 MERA}, 1 oper.}\\
\midrule
2	&	15-17	&  4	&	2  \\
3	&	21-27	&  8	&	3  \\
4	&	27-37	&  16	&	8   \\
5	&	33-47	&  32	&	21   \\
6	&	39-57	&  64	&	50   \\
7	&	45-67	&  128	&	111   \\
8	&	51-77	&  256	&	236   \\
9	&	57-87	&  512	&	489   \\
10	&	63-97	&  1024	&	998   \\
\midrule
\multicolumn{4}{l}{\textbf{1D 2:1 MERA}, 2 oper.}\\
\midrule
2	&	16-22	&  3	&	2   \\
3	&	22-38	&  7	&	4   \\
4	&	28-54	&  15	&	9   \\
5	&	34-70	&  31	&	21   \\
\boxitblue{1.47in} 6	&	40-86	&  63	&	48   \\
7	&	46-102	&  127	&	106  \\
8	&	52-118	&  255	&	227   \\
9	&	58-134	&  511	&	475   \\
10  &   64-150	&  1023	&	978   \\
\midrule
\multicolumn{4}{l}{\textbf{2D 4:1 MERA}, 1 oper.}\\
\midrule
2	&	23-25	&	12	&	2   \\
3	&	29-55	&	68	&	7   \\
4	&	43-81	&	256	&	122   \\
\midrule
\multicolumn{4}{l}{\textbf{2D 4:1 MERA}, 2 oper.}\\
\midrule
2	&	29-38	&	9	&	3   \\
3	&	40-84	&	88	&	40   \\
\boxitred{1.47in} 4	&	81-132	&	223	&	134   \\
\bottomrule
\end{tabular}

\phantom{00}
\phantom{00}
\phantom{00}
\phantom{00}
\phantom{00}
\phantom{00}
\phantom{00}
\phantom{00}
\phantom{00}
\phantom{00}
\phantom{00}
\phantom{00}
\begin{myboxred}
\footnotesize
\centering
\begin{tabular}{rrrr}
\multicolumn{4}{c}{\textbf{2D 4:1 MERA}}\\
\toprule
\multicolumn{4}{c}{\textbf{$\ell = 4$, 2 oper.}}\\
\midrule
$|V|$ & $|E|$ & $|{\mathcal{S}}|$ & $|\widehat{\mathcal{S}}|$ \\
\midrule
88	&	259	&	4	&	1	\\
90	&	264	&	4	&	1	\\
104	&	305	&	16	&	9	\\
108	&	315	&	8	&	1	\\
110	&	320	&	4	&	1	\\
110	&	323	&	16	&	9	\\
112	&	331	&	8	&	5	\\
114	&	333	&	16	&	9	\\
116	&	341	&	16	&	13	\\
118	&	346	&	4	&	1	\\
120	&	351	&	28	&	25	\\
120	&	354	&	20	&	13	\\
122	&	359	&	8	&	1	\\
122	&	362	&	8	&	5	\\
124	&	367	&	4	&	1	\\
126	&	372	&	40	&	33	\\
130	&	385	&	8	&	1	\\
132	&	390	&	4	&	1	\\
132	&	393	&	6	&	3	\\
\bottomrule
\end{tabular}
\end{myboxred}
\end{multicols}

\caption{Summary of extended MERA benchmark data. (Center) For each lattice type (1D binary or 2D 4-ary) and number of operators, as the number of levels $\ell$ varies, we report the number of vertices $|V|$ in the resulting networks, the total number of networks $|\mathcal{M}|$, and the number of unique networks up to isomorphism $|\widehat{\mathcal{M}}|$. (Left, Right) These detailed tables each expand a row from the summary table, specifying the number of networks produced (total $|\mathcal{S}|$ and up to isomorphism $|\widehat{\mathcal{S}}|$), for each pair of values for the number of vertices $|V|$ and number of edges $|E|$. Note that sum of the $|\mathcal{S}|$'s in a detailed table sums to the corresponding $|\mathcal{M}|$ in the summary table, and likewise for $|\widehat{\mathcal{S}}|$ and $|\widehat{\mathcal{M}}|$.}
\label{table:mera_full}
\end{table}

Notably, given a fixed set of parameters needed to generator the MERA network, several choices of operator inputs can result in isomorphic (i.e., identical up to relabeling) networks.
As seen in Fig.~\ref{figure:mera}, placing one operator at various sites can result in both isomorphic and non-isomorphic networks, so care must be taken to generate an appropriate set of representative graphs.

\begin{figure}[!hb]
\begin{adjustwidth}{-2.25in}{0in}
\centering
\includegraphics[width=\linewidth]{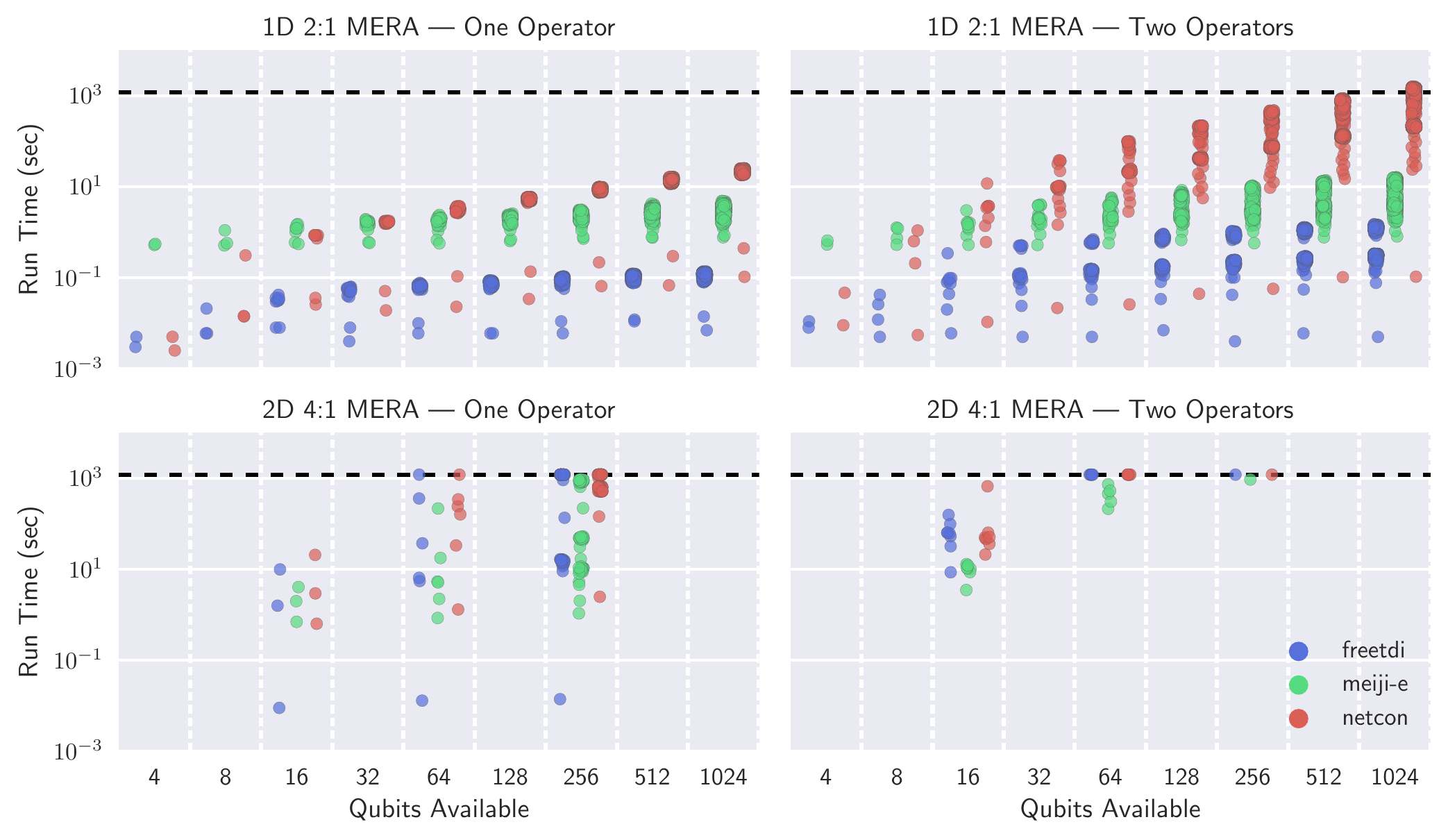}
\caption{Run times for the contraction sequence algorithms on select MERA networks, binned by number of qubits possible (number of $\mathcal{L}_0$ sites). All algorithms are timed out at 20 minutes (horizontal dashed line), and a network that remained unsolved by every algorithm is not included. 2D MERA with one operator had 48 of 131 networks that did not finish, and the two operator networks had 193 of 207 that did not finish.}
\label{figure:mera_qubit_runtime}
\end{adjustwidth}
\end{figure}

In our extended comparison, we generate 1D binary MERA and 2D 4:1 MERA with 1 and 2 operator placements.
For 1 operator, we generate a MERA network for every possible operator placement, then compute the unique networks up to isomorphism.
For 2 operators, we fix an operator at the first position (0 for 1D and (0,0) for 2D), then range the second operator over all possible positions; again we extract the unique graphs up to isomorphism.
Table~\ref{table:mera_full} summarizes the number of networks up to isomorphism and details some of their nuances.
For example, for a 1D 2:1 MERA with six coarsening levels and two operator placements, the network will have between 40 and 86 vertices based on operator placement (see the dark-blue highlighted row in the summary table). In this case, over all placements, $|\mathcal{M}| = 63$  networks were generated, but only $|\widehat{\mathcal{M}}| = 48$ were unique.
Looking into these unique networks even further (left table), we find that only one network had 40 vertices, and over half of the unique graphs had over 80 vertices.
While these networks may vary in the extreme cases, on average these networks are fairly predictable.

\begin{figure}[!b]
\begin{adjustwidth}{-2.25in}{0in}
\centering
\includegraphics[width=\linewidth]{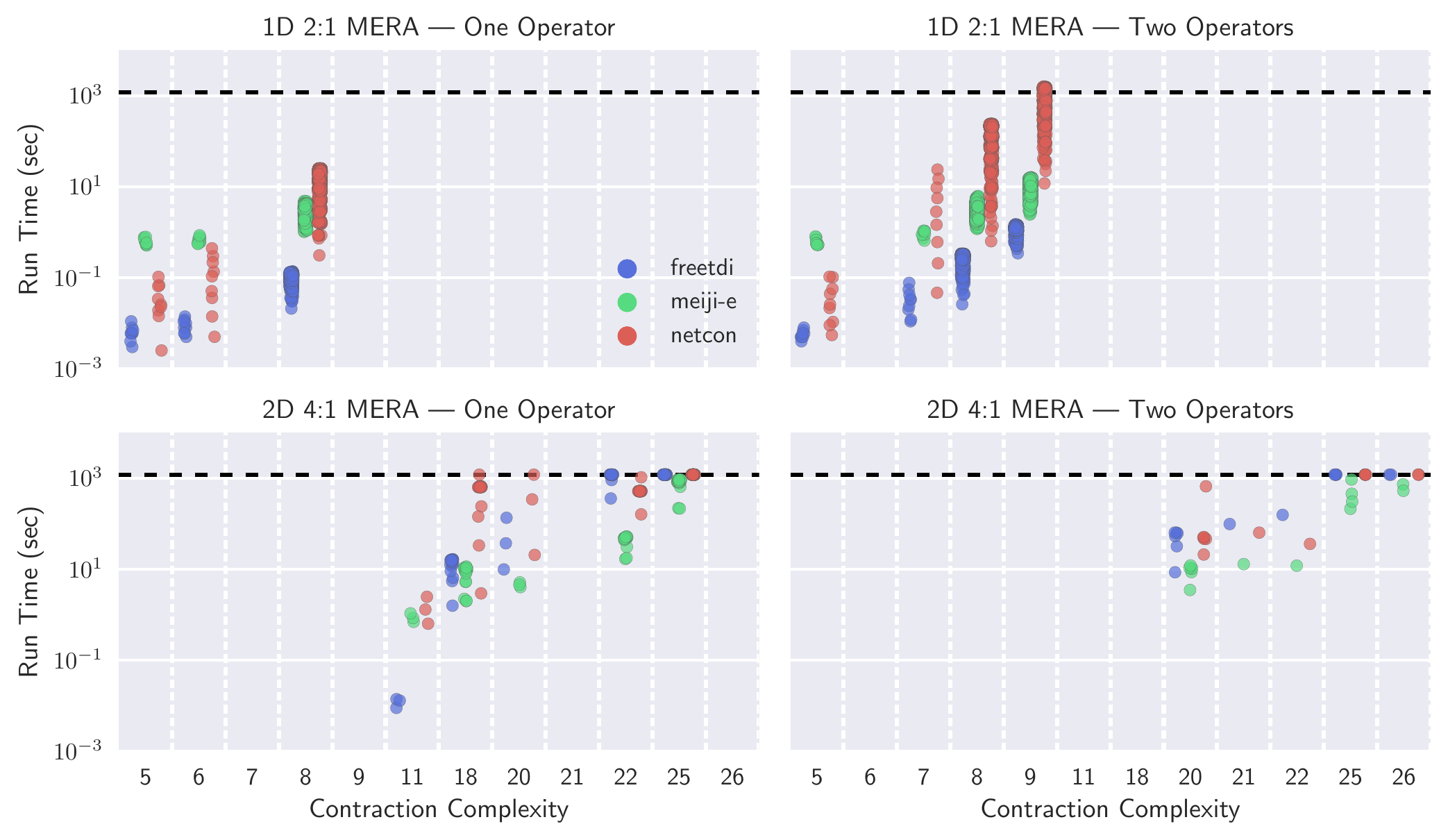}
\caption{Run times for the three algorithms on select MERA networks, binned by optimal contraction complexity. All algorithms are timed out at 20 minutes (horizontal dashed line), and a network that remained unsolved by every algorithm is not included. 2D MERA with one operator had 48 of 131 networks that did not finish, and the two operator networks had 193 of 207 that did not finish.}
\label{figure:mera_treewidth_comparison}
\end{adjustwidth}
\end{figure}

Fig~\ref{figure:mera_qubit_runtime} and Fig~\ref{figure:mera_treewidth_comparison} visualize the results of our extended experiment running contraction sequences algorithms on larger MERA networks using \software.
In Fig~\ref{figure:mera_qubit_runtime}, networks are binned by the number of qubits they can support for a quantum simulation (which is determined by the number of possible operator sites) and run time is reported in seconds.
For the networks that were included, a timeout of 20 minutes was used; if no algorithm could solve an instance within 20 minutes then the network was not included as a datapoint.
From this perspective, \netcon appears slower than both treewidth-based solvers, and \freetdi generally dominates \meijie in 1D MERAs.
However, on 2D MERAs with one operator, all algorithms seem roughly equal for 16-qubits and 64-qubits, with \meijie pulling slightly ahead on 256-qubits.
The improved scaling of \meijie is reflected in 2D MERAs with two operators, although here we begin to hit the limit of exact algorithms with a 20 minute timeout.

In Fig~\ref{figure:mera_treewidth_comparison}, networks are binned based on their contraction complexity, then run times are reported for each algorithm.
From this perspective we find a stipulation for \freetdi's dominance on 1D MERAs: the contraction complexity never exceeded 9.
Indeed, \freetdi performed well on 2D MERAs until the optimal contraction complexity reaches 18, at which point \meijie starts to outperform both algorithms.
These experiments depict that finding the optimal contraction sequence within 20 minutes becomes problematic when the contraction complexity reaches the mid-twenties, which may be useful for predicting when heuristics should be used.

We note that while the experiment from \cite{pfeifer2014faster} contained a graph with contraction complexity of 26, this network only had 27 tensors and 55 edges, whereas our 2D 4:1 MERA networks with contraction complexity 26 could have as many as 393 edges.
This fact implies that search-space--based approaches need pruning techniques that prune exponentially-many sequences in the number of network edges, otherwise the run time will scale without the optimal contraction complexity necessarily increasing.

\section*{\qtorchsection}
\label{section:application-qtorch}
In this section we compare exact treewidth solvers from PACE with the \quickbb solver used in Fried et al.'s qTorch tensor network simulator~\cite{fried2017qtorch}.
This comparison is run on quantum circuits constructed with the QAOA method for solving \maxcut on $r$-regular graphs.
We find that exact contraction sequences on these graphs can be computed in less time than needed to execute the tensor contractions, allowing us to discuss the total simulation time without using an untimed preprocessing step.
Rerunning the comparison with Microsoft's \liquid simulator from~\cite{fried2017qtorch}, we find that tensor networks are competitive against state-of-the-art simulators, allowing speedups when little information is needed to represent the network (i.e., on sparse graphs), and a potential solution for the 22-qubit limit currently built into \liquid.

\begin{figure}[!ht]
\centering
\includegraphics[width=0.75\linewidth]{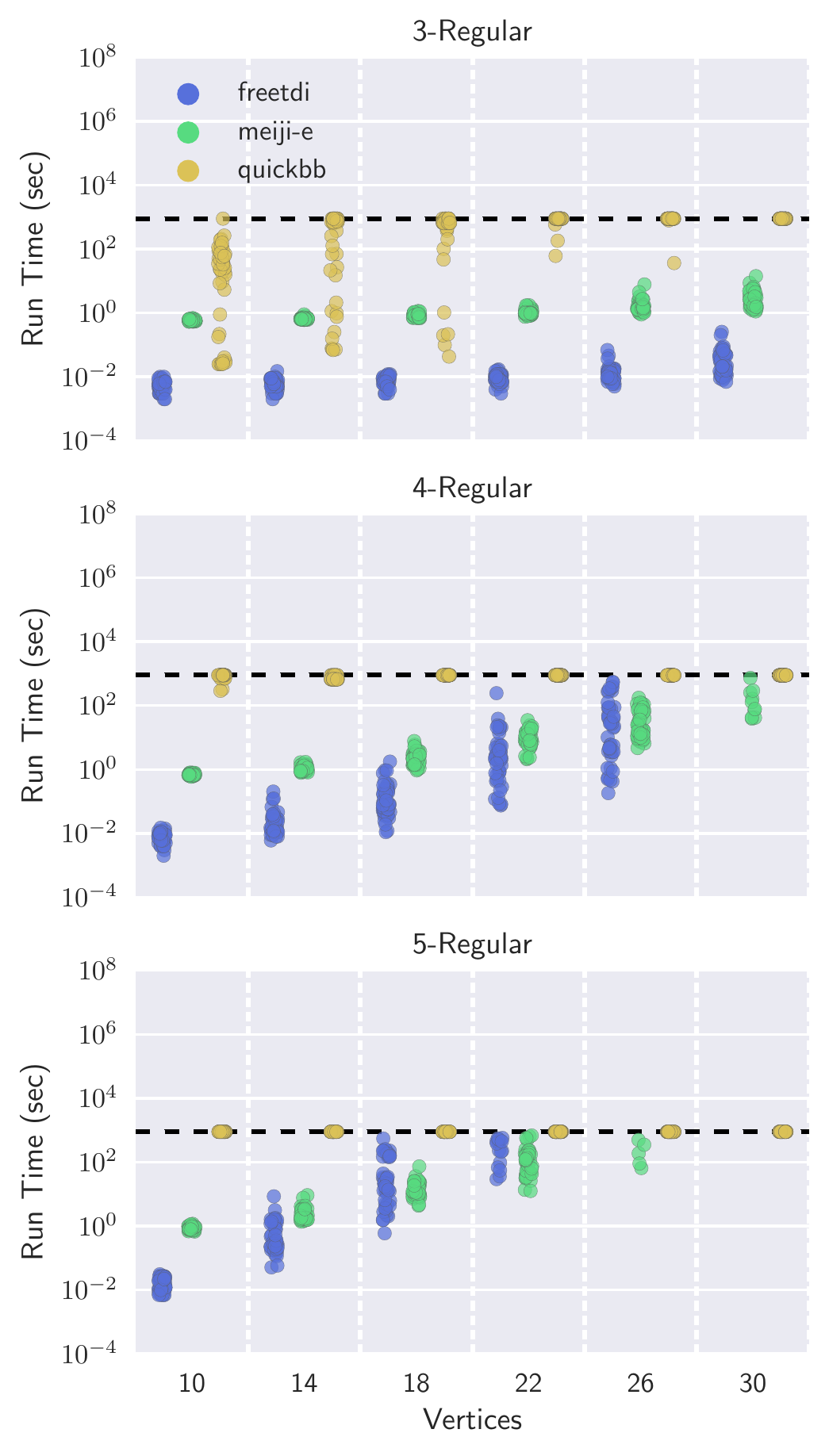}
\caption{Run times for \freetdi, \meijie, and \quickbb on QAOA circuits for computing \textsc{MaxCut} on $r$-regular graphs. 25 random regular graphs are generated at each $r, |V|$ level using NetworkX, and algorithms were timed out at 15 minutes (horizontal line).}
\label{figure:qtorch_treewidth_times}
\end{figure}

\begin{table}[!ht]
\centering
\begin{tabular}{llrrrrrr}
\toprule
\multicolumn{2}{c}{\textbf{Network}}& \multicolumn{6}{c}{\textbf{Optimal Contraction Complexity}} \\
\cmidrule(lr){1-2}
\cmidrule(lr){3-8}
$r$ & $|V|$ & \textbf{Samples} & \textbf{Mean} & \textbf{S.D.} & \textbf{Min} & \textbf{50\%} & \textbf{Max} \\
\midrule
3 & 10 &     25 &   5.0 &  0.6 &    4 &    5 &    6 \\
  & 14 &     25 &   5.2 &  0.6 &    4 &    5 &    6 \\
  & 18 &     25 &   6.0 &  0.8 &    5 &    6 &    7 \\
  & 22 &     25 &   6.3 &  0.9 &    5 &    6 &    8 \\
  & 26 &     25 &   6.7 &  0.7 &    6 &    7 &    8 \\
  & 30 &     25 &   7.8 &  0.6 &    7 &    8 &    9 \\
\midrule
4 & 10 &     25 &   6.5 &  0.5 &    6 &    7 &    7 \\
  & 14 &     25 &   7.7 &  0.8 &    6 &    8 &    9 \\
  & 18 &     25 &   8.7 &  0.7 &    8 &    9 &   10 \\
  & 22 &     25 &  10.2 &  0.8 &    8 &   10 &   11 \\
  & 26 &     24 &  11.2 &  1.2 &    9 &   12 &   13 \\
  & 30 &      5 &  12.0 &  0.7 &   11 &   12 &   13 \\
\midrule
5 & 10 &     25 &   7.7 &  0.6 &    6 &    8 &    9 \\
  & 14 &     25 &   9.6 &  0.7 &    8 &   10 &   11 \\
  & 18 &     25 &  11.3 &  0.7 &   10 &   11 &   13 \\
  & 22 &     23 &  12.7 &  0.8 &   11 &   13 &   14 \\
  & 26 &      1 &  12.0 &  NaN &   12 &   12 &   12 \\
\bottomrule
\end{tabular}

\caption{Exact contraction complexities found using \freetdi and \meijie on QAOA circuits for computing \textsc{MaxCut} on $r$-regular graphs. 25 random regular graphs are generated at each $r, |V|$ level using NetworkX, and algorithms were timed out at 15 minutes. Timed out values were dropped from the data, resulting in less than 25 Samples for some parameter values.}
\label{table:qtorch_treewidths}
\end{table}

\subsection*{Computing Contraction Complexity}
Before comparing total simulation times, we start by evaluating the times required to find optimal contraction sequences. Because non-optimal contraction sequences lead to exponentially-slower downstream simulation times, we are particularly interested in exploring the limits of exact solvers.

The data for this experiment comes from qTorch's QAOA quantum circuit constructor, which computes \maxcut on a specified (arbitrary) graph.
Reproducing circuits similar to the original qTorch experiments, we use $r$-regular graphs for $r \in \{3, 4, 5\}$, generated with NetworkX and seeded for easy replicability.

Fig~\ref{figure:qtorch_treewidth_times} visualizes the results from using 25 random graphs for each $(r, |V|)$ pair and a 15 minute timeout; Table~\ref{table:qtorch_treewidths} provides further information relating $(r, |V|)$ pairs with their optimal contraction complexity.
Similar to the sparse MERA graphs, we find that \freetdi dominates run times and is up to five orders of magnitude faster than \quickbb on the smallest graphs. \meijie again scales better than \freetdi, allowing it to compute optimal contraction sequences for several networks within the timeout that other algorithms could not finish.
Notably, \quickbb is an anytime algorithm that finds increasingly better perfect elimination orderings, so it always provides a (potentially non-optimal) solution when given a timeout. Similar behavior may be adapted from heuristic versions of PACE submissions (e.g., \meijie), but this functionality is left as future work.

\subsection*{Simulation Run Times}

Computing downstream simulation times using qTorch, we first evaluate how simulation time correlates with contraction complexity. This comparison enables us to quantify how downstream runtime is impacted by the contraction complexity in practice (theoretical analysis predicts a exponential increase due to the size of the merged tensors, but it is possible that some downstream code mitigates this impact). Further, it provides for a finer-grained comparison of the quality of contraction sequences produced by \freetdi and \meijie, as measured by resulting simulation run time.
Although both algorithms produce sequences with the same contraction complexity (and thus have the same leading-order term in the simulation time complexity), this term may occur during the contraction sequence between 1 and $|V|$ times.
By examining raw simulation times, we may infer that one algorithm tends towards `higher quality' contraction sequences than another.

\begin{figure}[!ht]
\centering
\includegraphics[width=\linewidth]{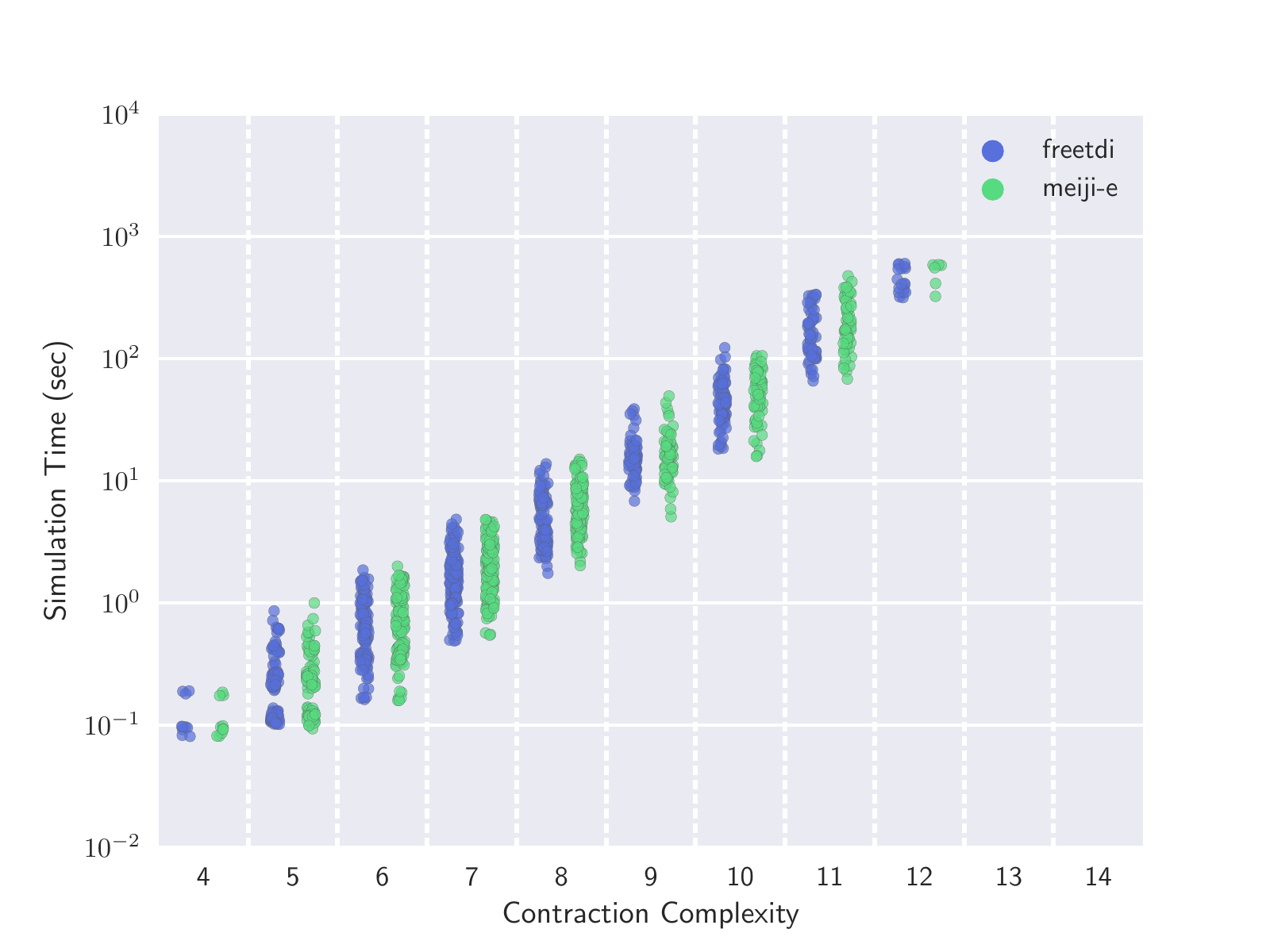}
\caption{Simulation time is tightly correlated with the contraction complexity of a network. While exact algorithms \freetdi and \meijie may generate different tree decompositions and thus contraction sequences with the same treewidth, the differences have little impact on simulation times.}
\label{figure:qtorch_run_time_vs_treewidth}
\end{figure}

As shown in Fig~\ref{figure:qtorch_run_time_vs_treewidth}, simulation time indeed has an exponential dependence on the contraction complexity.
Additionally, the contraction sequences produced by \freetdi and \meijie appear to be of comparable quality over the same corpus of networks.
We observe that the range of run times for a given contraction complexity is nearly always an order of magnitude, meaning that the variance in run time scales proportional to the total run time.
This observation may be useful for reliably predicting simulation time based on a network's known contraction complexity, which may be useful for optimizing future simulators.

In our final experiment, we compare qTorch paired with optimal contraction sequences to Microsoft's \liquid simulator. Refer to the \software section for workstation specifications. Previous comparisons to \liquid included non-optimal contraction sequences found by \quickbb, which may have caused the downstream simulation to be exponentially more expensive. Additionally, the experiments in~\cite{fried2017qtorch} were structured so that \quickbb was run for 3000 seconds per network ahead of time, which resulted in impractical total run times for lower levels of regularity.

\begin{figure}[!ht]
\centering
\includegraphics[width=0.75\linewidth]{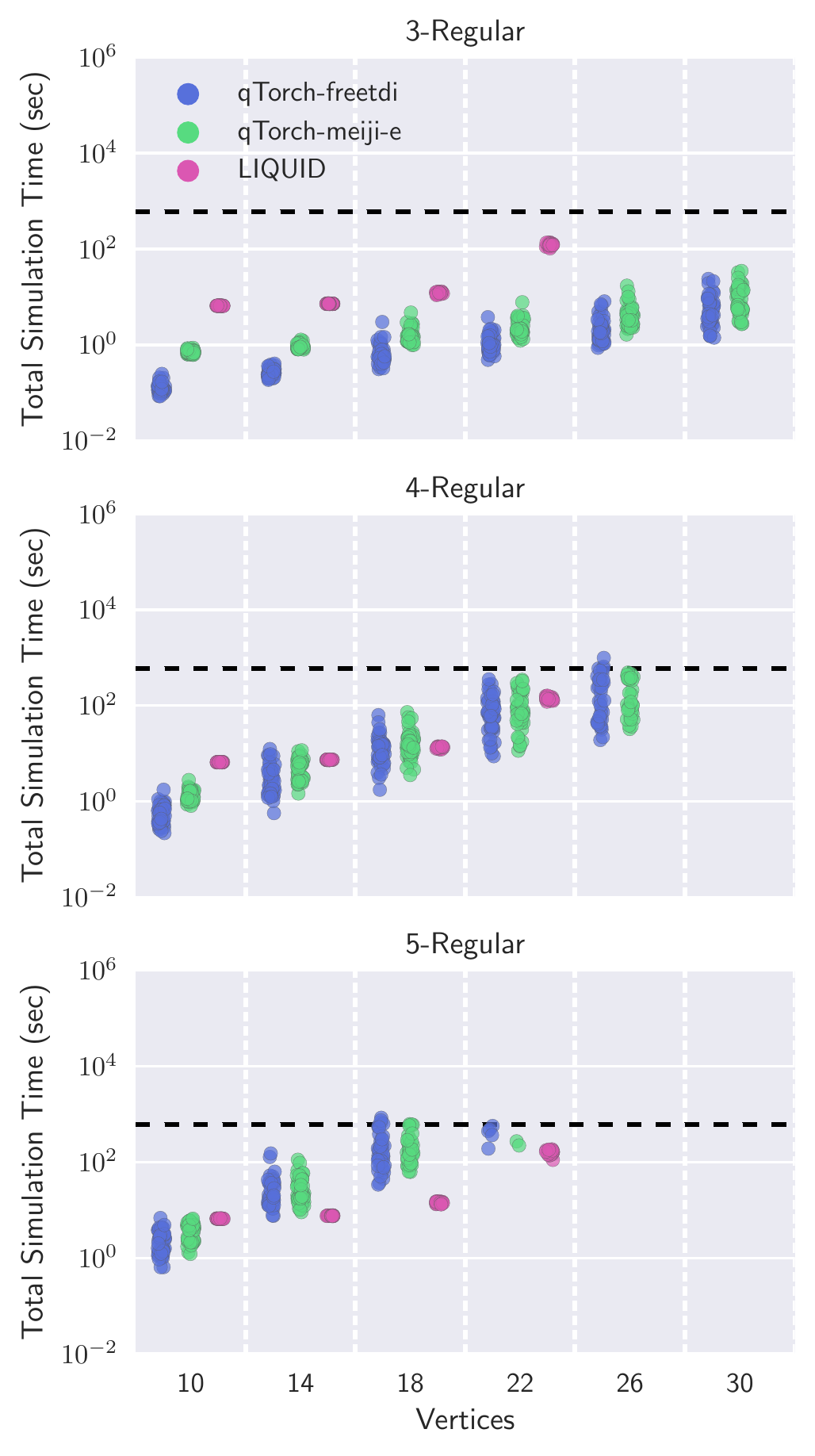}
\caption{Simulation times of the qTorch tensor network simulator \cite{fried2017qtorch} with contraction sequences produced by exact treewidth algorithms vs. Microsoft's LIQUID solver \cite{wecker2014liquid}. Total simulation time includes both computation of the contraction sequence using \software and tensor network simulation time using qTorch. A timeout of 900 seconds is used for computing the contraction sequence (horizontal dashed line), and a simulation is not run unless an optimal contraction sequence is found by at least one contraction sequence algorithm. LIQUID is limited to simulations up to 22 qubits.}
\label{figure:liquid_comparison}
\end{figure}

Results for this experiment are visualized in Fig~\ref{figure:liquid_comparison}.
Aligning with previous conclusions in \cite{fried2017qtorch}, both 3-regular and 4-regular graphs are more quickly simulated on tensor networks than \liquid.
We additionally find that tensor networks can scale beyond the 22-qubit limit imposed on \liquid (supposedly for exponential system memory usage).
Even with 5-regular graphs, qTorch remains competitive as an alternative simulator.

Comparing Fig~\ref{figure:qtorch_treewidth_times} with Fig~\ref{figure:liquid_comparison}, it is also clear that computing the contraction complexity is no longer the primary bottleneck.
Instead, efforts should now be directed to improving the contraction simulation times.

\section*{Conclusion}
\label{section:conclusion}
In summary, \software provides an open source, extendable platform for comparing contraction sequence algorithms for tensor networks.
By packaging conversion utilities with containerized solvers, we remove both the theoretical and engineering difficulties preventing practitioners from running any contraction sequence solver on any tensor network.
Additionally, we demonstrate the framework's applicability by reproducing and significantly extending several prior empirical evaluations.
With MERA networks, we introduce a more extensive and difficult benchmark dataset which allows identification of solutions that will scale (e.g., \freetdi on 1D MERA networks), subtler performance differences between algorithms (e.g., how \meijie scales better for larger contraction complexities), and areas where new approaches are needed (e.g, 2D, 2-operator MERA, where 193 of 207 networks timed out).
With qTorch on QAOA data, we were able to validate the exponential dependence of run time on optimal contraction complexity, and produce a total simulation time more representative of the full pipeline required for simulation.
In doing so, we illuminate the urgent need for improved contraction times when the contraction sequence is known.

For contraction sequence algorithms, several avenues offer promising future work.
With large MERA networks we found that exact solvers for optimal contraction sequences had prohibitively high run times, a difficulty which may require using non-exact heuristics.
Several treewidth-based algorithms have heuristic formulations in a different track of the PACE 2017 challenge~\cite{dell2017pace}, and our comparison against domain-specific algorithms suggests that these PACE submissions would be the natural starting point for an investigation into heuristics. Additionally, the use of heuristics involves a trade-off between finding a sequence with smaller complexity and the additional search time, which is not a problem for exact solvers. Exploring this trade-off in practical applications may be of interest.

Another consideration is that tensor network simulations are increasingly run on high-performance computing (HPC) systems.
Modern HPC platforms scale-up performance with parallelism and heterogenous computing accelerators (such as GPUs and FPGAs).
As seen in recent work~\cite{van2017computing}, adapting treewidth's state-of-the-art dynamic programming algorithms to these platforms is a non-trivial task, but would be of significant potential impact to the quantum computing community.

\section*{Acknowledgments}
\label{section:acknowledgments}
Eugene Dumitrescu acknowledges support from the Laboratory Directed Research and Development Program of Oak Ridge National Laboratory, managed by UT-Battelle, LLC, for the U.S. Department of Energy.
Allison Fisher, Blair D. Sullivan, and Andrew Wright were supported in part by the Gordon \& Betty Moore Foundation's Data-Driven Discovery Initiative through Grant GBMF4560 to Blair D. Sullivan, and
the NC State Provost's Professional Experience Program.
Timothy D. Goodrich was funded with Government support under and awarded by DoD, Air Force Office of Scientific Research, National Defense Science and Engineering Graduate (NDSEG) Fellowship, 32 CFR 168a.
Travis Humble acknowledges support from the U.S. Department of Energy, Office of Science Advanced Scientific Computing Research and Early Career Research programs.

\bibliography{references.bib}
\end{document}